\documentclass[referee,sn-basic]{sn-jnl}

\usepackage{graphicx}%
\usepackage{multirow}%
\usepackage{amsmath,amssymb,amsfonts}%
\usepackage{amsthm}%
\usepackage{mathrsfs}%
\usepackage[title]{appendix}%
\usepackage{xcolor}%
\usepackage{textcomp}%
\usepackage{manyfoot}%
\usepackage{booktabs}%
\usepackage{algorithm}%
\usepackage{algorithmicx}%
\usepackage{algpseudocode}%
\usepackage{listings}%
\usepackage{orcidlink}%

\theoremstyle{thmstyleone}%
\theoremstyle{thmstyletwo}%

\theoremstyle{thmstylethree}%
\newcommand{\code}[1]{\texttt{#1}}%
\newcommand{\pkg}[1]{\textbf{\texttt{#1}}}%
\newcommand{\class}[1]{\code{#1}}%

\begin{document}

\title[Power Analysis of Exposure Mixture Studies via Monte Carlo Simulations]{Power Analysis of Exposure Mixture Studies via Monte Carlo Simulations}


\author*[1]{\fnm{Phuc} \sfx{H} \sur{Nguyen} \orcidlink{0000-0002-6206-0194}}
\email{phuc.nguyen.rcran@gmail.com}

\author[1]{\fnm{Amy} \sfx{H} \sur{Herring}}

\author[2]{\fnm{Stephanie} \sfx{M} \sur{Engel}}

\affil[1]{
\orgdiv{Department of Statistical Science}, 
\orgname{Duke University}, 
\orgaddress{\city{Durham}, \postcode{27710}, \state{NC}, \country{USA}}}

\affil[2]{
\orgdiv{Department of Epidemiology}, 
\orgname{The University of North Carolina at Chapel Hill}, 
\orgaddress{\city{Chapel Hill}, \postcode{27516}, \state{NC}, \country{USA}}}


\abstract{Estimating sample size and statistical power is an essential part of a good epidemiological study design. Closed-form formulas exist for simple hypothesis tests but not for advanced statistical methods designed for exposure mixture studies. Estimating power with Monte Carlo simulations is flexible and applicable to these methods. However, it is not straightforward to code a simulation for non-experienced programmers and is often hard for a researcher to manually specify multivariate associations among exposure mixtures to set up a simulation. To simplify this process, we present the \texttt{R} package \pkg{mpower} for power analysis of observational studies of environmental exposure mixtures involving recently-developed mixtures analysis methods. The components within \pkg{mpower} are also versatile enough to accommodate any mixtures methods that will developed in the future. The package allows users to simulate realistic exposure data and mixed-typed covariates based on public data set such as the National Health and Nutrition Examination Survey or other existing data set from prior studies. Users can generate power curves to assess the trade-offs between sample size, effect size, and power of a design. This paper presents tutorials and examples of power analysis using \pkg{mpower}.}

\keywords{power analysis, environmental chemical mixtures, observational study, Monte Carlo simulation, \texttt{R} package}



\maketitle

\section{Introduction}\label{intro}

Researchers need to estimate sample size and statistical power as part of good study planning. The power of a test is the probability that it rejects the null hypothesis at a specific significance level under a prespecified alternative hypothesis. Calculating power requires assumptions about relevant quantities, such as variance, effect size, and sample size. For standard hypothesis tests, including one or two sample t-tests, one-way balanced ANOVA, and chi-squared tests, closed-form equations exist to calculate the power. However, in general, power formulas are not available for complex analysis methods, such as statistical models recently developed for mixtures analysis. In these situations, we can estimate power using Monte Carlo (MC) simulations, an approach that is flexible and applicable to a large class of statistical models \citep{gelman2007}. Each simulation involves first sampling data from a hypothesized data generative model and then fitting an inference model to the data. Out of a large number of simulations, the proportion of times in which a hypothesis test is rejected can be used to estimate power. Biomedical research has used simulations for power calculation in randomized trials \citep{oview}, longitudinal studies \citep{gastanaga2006power}, and observational studies with measurement error or drop out \citep{landau2013sample}, to name a few. To our knowledge, this is the first software developed specifically for power calculations in observational studies of chemical exposure mixtures, where controlling for several covariates and multiple correlated exposures is important.

Furthermore, power calculation should be based on the intended inference model, which aligns with the analysis goals. Exposure mixture studies are almost always observational in nature due to ethical constraints in exposing subjects to potentially harmful chemicals. Researchers then use regression methods to parse out associations between chemical mixtures and health outcomes of interest. Statistical models for mixtures aim to achieve several goals, including but not limited to: (1) identifying individual chemicals critical to the health outcome of interest; (2) estimating the health effects of individual (or combinations of) chemicals in the presence of multicollinearity; and (3) examining synergistic interactions between chemicals \citep{sun-review}. Several challenges are associated with these goals, and no single approach is best at addressing all of them. For instance, regularized multiple regression with main effects and interactions can identify critical chemicals, and examine synergistic effects, but tend to be limited to pairwise interactions \citep{bienlasso, limlasso}. Kernel-based regressions can account for high-order interactions and nonlinear effects but tend to not scale well with the number of chemicals or observations \citep{hamra-review}. Many models incorporate factor analysis for dimensionality reduction to deal with multicollinearity \citep{sun-review, mixselect, fin}. The choice of the model depends on the goal of analysis and characteristics of the data set \citep{sun-review}. Therefore, power analysis should be conducted for the specific method used in the analysis. MC simulation enables power calculation for different combinations of hypothesized exposure-response relationships and inference models.

We present the \texttt{R} package \pkg{mpower} containing building blocks to set up MC simulations for estimating power for observational studies of environmental exposure mixtures.
Some packages exist for conducting power analysis using MC simulations, including \pkg{simr} \citep{simr}, \pkg{skpr} \citep{skpr}, and \pkg{simglm} \citep{simglm}. However, existing packages are not tailored to exposure mixture studies or compatible with mixture inference methods. Package \pkg{skpr} \citep{skpr} provides options to simulate data from optimal factorial designs and split-plot designs, often not applicable to observational epidemiological studies. \cite{oview} provided example \texttt{R} code to do power analysis through MC simulations for complex cluster design, but it is unclear how to incorporate continuous exposures. Package \pkg{simglm} \citep{simglm} provides a general framework for power analysis for multi-leveled, mixed-scaled data but does not facilitate the simulation of correlated exposures. Package \pkg{simr} \citep{simr} conducts simulations based on models fitted on real-world data but only works with generalized linear mixed models.

Our package \pkg{mpower} allows users to simulate realistic exposure data and mixed-typed covariates based on existing data sets such as the National Health and Nutrition Examination Survey (NHANES). It enables users to easily conduct power analysis for recently-developed statistical methods for mixtures that lack closed-form power formulas, including Bayesian Kernel Machine Regression (BKMR) \citep{bkmr}, Bayesian Model Averaging (BMA) \citep{hoetingbma}, MixSelect (MS) \citep{mixselect}, Factor Analysis for Interactions (FIN) \citep{fin}, Bayesian Weighted Sums (BWS) \citep{bws}, and Quantile G-Computation (QGC) \citep{qgcomp}. These features enable researchers, even those not experienced with \texttt{R}, to quickly set up MC simulations for exposure mixture studies. Simultaneously, the building blocks within \pkg{mpower} are versatile enough to integrate with other \texttt{R} packages and user-defined functions, empowering advanced \texttt{R} users to work with statistical models for mixtures beyond those provided in \pkg{mpower}. Because simulation-based power analysis for complex models can be computationally intensive, we allow parallel computation when users have access to multi-core computer processors. In Section \ref{sec:method}, we describe algorithms for power analysis using MC simulations, simulation of correlated mixed-scaled covariates and exposures, automatic scaling of effect size, and state-of-the-art statistical methods for chemical mixtures. In Section \ref{sec:tutorial}, we provide a tutorial on the package's functionalities with example codes. In Section \ref{sec:example}, we provide a tutorial on end-to-end power analyses with synthetic data and NHANES, demonstrating that simulation is a reliable approach to estimating power and a potentially useful tool for inference model comparison.

\section[Methodology]{Methodology} \label{sec:method}

\subsection{Power analysis using Monte Carlo simulation}

We can use MC simulation to estimate power given a data generative model and an inference model. Let $s$ be the number of simulations, $n$ be the sample size of one simulation, $y_i, i=1,...,n$, be the outcome of the $ith$ subject, and $\boldsymbol{\theta} = (\boldsymbol{\theta}_Y, \boldsymbol{\theta}_X)$ be the parameters (i.e., intercepts, main effects, interactions, variance components, correlations among predictors, ...) in a data generative model that realistically represents the multivariate association among the predictors and a hypothesis about the relationships between the predictors and outcome. As a result, this data generative model involves a generative model for the predictors and a generative model for the outcome. The parameters $\boldsymbol{\theta}$ may be chosen based on prior knowledge of the associations among predictors, effect size or estimates from previous studies \citep{oview, simr}.

\begin{algorithm}
	\caption{Power analysis using Monte Carlo simulation}  
	\label{alg:power_mc}
	\begin{algorithmic}[1]
	    \State Set $\boldsymbol{\theta}$ based on prior knowledge and/or estimates from previous studies.
		\For {$iteration=1,2,\ldots,s$}
		    \State Simulate $n$ independent samples of the predictors $\boldsymbol{x}_i, i=1,...,n$ given $\boldsymbol{\theta}_X$.
		    \State Simulate $y_i$ given parameters $\boldsymbol{\theta}_Y$ and predictors $\boldsymbol{x}_i$.
			\State Perform analysis with an inference model and record if the effects meet some ``significance'' criterion (e.g. p-value less than $0.05$).
		\EndFor
	\State Calculate the power as the proportion of simulations meeting the defined criterion.
	\end{algorithmic} 
\end{algorithm}

\noindent We may repeat Algorithm \ref{alg:power_mc} with different values for $\boldsymbol{\theta}$ and $n$ to create a power curve. Power curves are useful to study the trade-offs between power, sample size, and effect size.

\subsection{Simulate correlated predictors}

We provide three methods to construct a generative model for correlated predictors. One approach to simulate samples from a realistic joint distribution of the predictors is to resample them from existing data. Another option is to estimate the joint distribution of the existing data and simulate from it. In this package, we use a semiparametric Bayesian Gaussian copula (SBGC) \citep{hoff2007sgc} to estimate the joint distribution of mixed binary, ordinal, and continuous data. This method is based on Sklar's Theorem that any multivariate probability distribution can be written in terms of its univariate marginal probability distributions and the copula. For example, let $x_1, ..., x_p$ be $p$ random variables with marginal univariate CDF's $F_1, ..., F_p$, not necessarily continuous, and an unknown joint distribution we want to estimate. The transformed variables $u_j=F_j(x_j)$'s all have uniform marginals. A copula models the joint distribution of $u_1,...,u_p$, which also describes the dependence amongst the $x_j$'s separately from $F_1,...,F_p$. We use a Gaussian copula which has the following form:

\begin{align}
    (z_1,...,z_p)^T|\boldsymbol{C} &\sim N_p(\boldsymbol{0}, \boldsymbol{C}) \\
    x_j &= F_j^{-1}[\Phi(z_j)], j=1,...,p
    \label{eqn:copula}
\end{align}

\noindent where $z_j$'s are latent variables, $\boldsymbol{C}$ is a correlation matrix of the latent variables, $F_j^{-1}$ is the pseudo-inverse of $F_j$, and $\Phi$ the CDF of a standard normal. Since the closed form of the $F_j$'s are often unknown in practice, we can estimate them using the empirical distributions from $n$ observations $\hat{F}_j(t) = \frac{1}{n}\sum_{i=1}^n \boldsymbol{1}_{x_{ij} \leq t}$. Hoff proposed using the extended rank likelihood to estimate $\boldsymbol{C}$ without depending on estimates of $F_j$'s \citep{hoff2007sgc}, which works well for mixed-scaled data, and implemented the Bayesian inference algorithm in the \pkg{R} package \pkg{sbgcop} \citep{sbgcop}. The uncertainty in the estimation can be propagated through the whole power calculation straightforwardly by sampling from the copula's parameters' posteriors before simulating new data. Additionally, no specification of prior distributions or parametric form of the univariate marginal distributions is necessary, ensuring ease of use. We summarize how to simulate data from a generative model defined by a Gaussian copula in Algorithm \ref{alg:copula}.

\begin{algorithm}
	\caption{Simulate $p$ correlated predictors from a Gaussian copula} 
	\label{alg:copula}
	\begin{algorithmic}[1]
	    \State Given the univariate marginal CDF $F_j$'s defined by users or estimated with the empirical CDF.
		\For {$i=1,2,\ldots,n$}
		    \State Sample a correlation matrix $\boldsymbol{C}$ from the SBGC posterior (function\code{sbgcop::sbgcop.mcmc()}) or from Algorithm \ref{alg:alg_rand_cor}. 
			\State Sample the latent variables $\boldsymbol{z}_i \sim N_p(0, \boldsymbol{C})$.
			\State Get the predictors $\boldsymbol{x}_i$ through the transformation $x_j = F_j^{-1}[\Phi(z_j)], j=1,...,p$.
		\EndFor
		\State Return an $n \times p$ matrix $X$.
	\end{algorithmic} 
\end{algorithm}

Alternatively, if we have no existing data to estimate the correlation matrix $\boldsymbol{C}$ of the latent variables and the marginal CDF's $F_j$ of the observed variables, we can specify them manually. For instance, we may assume a continuous predictor to be normally distributed with a given mean and standard deviation, or an ordinal predictor to be from a multinomial distribution with given probabilities. As for the correlation matrix $\boldsymbol{C}$, given reasonable guesses on pairwise correlations between predictors, we can sample positive definite matrices with desirable structures using Algorithm \ref{alg:alg_rand_cor} from \citep{cvine}. This can add additional uncertainty into the samples of hypothetical data and ensure that $\boldsymbol{C}$ is a valid correlation matrix. Let $\rho_{ij;kL}$ be the partial correlation of the $i^{th}$ and $j^{th}$ variables while holding the $k^{th}$ variable and variables in set $L$ constant, where $L$ contains indices distinct from $\{i,j,k\}$. We can calculate partial correlation $\rho_{ij;L}$ using the following formula from \citep{cvine}:

\begin{equation}
    \rho_{ij;L} = \rho_{ij;kL}
    \sqrt{(1-\rho_{ik;L}^2)(1-\rho_{jk;L}^2)} + \rho_{ik;L}\rho_{jk;L}
    \label{eqn:partial-r-formula}
\end{equation}

The partial correlation $\rho_{ij}$ where $L$ is the empty set is equivalent to the correlation $r_{ij}$ between the $i^{th}$ and $j^{th}$ variables. Equation~\ref{eqn:partial-r-formula} shows that a $p \times p$ random correlation matrix can be parametrized in terms of pairwise correlations $\rho_{12},...,\rho_{1p}$ and partial correlations $\rho_{23;1}$,...,$\rho_{2p;1}$,$\rho_{34;12}$,..., $\rho_{3p;12}$,...,$\rho_{(p-1)p;1...p-2}$. These choices of partial correlations correspond to a C-vine \citep{bedford2002vines}. They can independently take values on the interval $(-1,1)$ following any density with the right support \citep{cvine}. We sample them from independent $Be(\alpha, \beta)$ distributions (scaled to be on $(-1, 1)$) with appropriate hyperparameters to induce a correlation matrix with desirable structures. Applying Equation~\ref{eqn:partial-r-formula} iteratively to smaller index sets $L$, we can calculate every element of the correlation matrix from these partial correlations (Algorithm \ref{alg:alg_rand_cor}). For example, $\rho_{23} = \rho_{23;1} \sqrt{(1-\rho_{12}^2)(1-\rho_{13}^2)} + \rho_{12}\rho_{13}$ where $k=1, L = \emptyset$, and $\rho_{23;1}, \rho_{12}, \rho_{13}$ are Beta random variables. More details can be found in \citep{joe, cvine}. When choosing $\{\alpha, \beta\}$, take into consideration that the mean of a Beta random variable is $\frac{\alpha}{\alpha + \beta}$, and the larger $\alpha$ or $\beta$ is, the smaller the variability of the samples. To automatically set reasonable parameters $\{\alpha_{ij}, \beta_{ij}\}$ for $\rho_{ij;1...i-1}$, we use the following heuristics given guesses of pairwise correlations, which are easier to specify than partial correlations. First, we calculate a partial correlation $\hat{\rho}_{ij;1...i-1}$ using the conditional multivariate normal formula \citep{eaton1983multivariate} from the rough guesses of pairwise correlations. If the guessed correlation matrix is singular, the generalized inverse can be used in the conditional normal formula. We then set $\alpha_{ij} = m \hat{\rho}_{ij;1...i-1}$ and $\beta_{ij} = m - \alpha_{ij}$ where $m$ controls the magnitude of the parameters. We find the heuristics to work well when the guessed pairwise correlations are not too extreme (e.g. 0.99). Simply put, we can turn rough guesses of pairwise correlations into a valid sampling distribution for the predictors.

\begin{algorithm}
	\caption{Sample random correlation matrix with C-vine} 
	\label{alg:alg_rand_cor}
	\begin{algorithmic}[1]
	    \State Set hyperparmeters $\alpha, \beta$ appropriately for each partial correlation
		\For {$i=1,2,\ldots,p-1$}
		    \For {$j=i+1,...,p$}
		        \State Sample partial correlation $\rho_{ij;1,...,i-1} \sim Be(\alpha_{ij}, \beta_{ij})$ on $(-1,1)$.
			    \State Calculate the correlation $r_{ij} = r_{ji} = \rho_{i,j}$ recursively using Equation~\ref{eqn:partial-r-formula}.
			\EndFor
		\EndFor
		\State Return the correlation matrix $\textbf{R}$ where $\textbf{R}_{ij} = r_{ij}$.
	\end{algorithmic}
\end{algorithm}

\subsection{A general notion of effect size}

It is not straightforward to set an effect size for nonlinear or complex mean functions. We propose using the signal-to-noise ratio (SNR) as a general proxy to the joint effect size for any mean function. The SNR can be viewed as an estimate for the ratio between the variation in the outcome explained by the true predictors and the variation due to noise. For the model $Y = f(X) + \epsilon$ with independent, additive Gaussian noise $\epsilon \sim N(0, \sigma_\epsilon^2)$, it is defined as:

\begin{equation}
    \text{SNR} = \frac{Var(f(X))}{\sigma_\epsilon^2}
    \label{eqn:snr-continuous}
\end{equation}

Fixing either the mean function variance or the noise variance, it is straightforward to scale the other so that a SNR $\gamma$ is achieved. We can estimate the variance of any mean function with $\hat{\sigma}_f^2 = \frac{1}{m-1}\sum_{i=1}^m (f(x_i) - \frac{1}{m}\sum f(x_i))^2$, where $x_1,...,x_m$ are i.i.d. realizations of predictors $X$ for a large number $m$. The specific calculations are summarized in Algorithm \ref{alg:scale_noise} and \ref{alg:scale_mean}. We may extend this measure of effect size to binary or count outcomes using the SNR estimator for generalized linear models (GLM) proposed by \citep{czanner2015snr}. The estimator is defined as:

\begin{equation}
    \widehat{\text{SNR}} = \frac{Dev(\textbf{w}, \bar{w}1) - Dev(\textbf{w}, \textbf{h}) + 1 - p}{Dev(\textbf{w}, \textbf{h}) + p}
    \label{eqn:snr-discrete}
\end{equation}

\noindent where $\textbf{w}, \textbf{h}$ are vectors of realizations of outcome $W$ and mean model $h(X)$ respectively, and $\bar{w} = \frac{1}{n}\sum_{i=1}^n w_i$. The residual deviance $Dev( \textbf{w}, \textbf{h})$ is a measure of discrepancy between the observed and predicted data, with the following forms for binary and count outcome \citep{mccullagh1983glm}:

\begin{align}
    \text{Bernoulli} &: 2\sum_{i=1}^n w_i log(\frac{w_i}{h_i}) + (1-w_i)log(\frac{1-w_i}{1-h_i}) \\
    \text{Poisson} &: 2 \sum_{i=1}^n y_i log(\frac{w_i}{h_i}) - (w_i - h_i)
    \label{eqn:deviance-discrete}
\end{align}

For binary or count outcomes, we can adjust parameters in the mean model $h$ until we get a desirable estimated $\widehat{\text{SNR}}$.

\begin{algorithm}
    \caption{Scale the noise variance}
    \label{alg:scale_noise}
    \begin{algorithmic}[1]
    \State Given a model of the outcome $Y = f(X) + \epsilon$ with independent, additive Gaussian noise $\epsilon \sim N(0, \sigma_\epsilon^2)$, the predictors $X \sim N(0,\Sigma_x)$, and the desired SNR $\gamma$.
    \State Sample $x_i \sim N(0,\Sigma_x), i=1,...,m$ for large $m$.
    \State Estimate $\hat{\sigma}_f^2 = \frac{1}{m-1}\sum_{i=1}^m (f(x_i) - \frac{1}{m}\sum f(x_i))^2$.
    \State Set the noise variance to $\sigma^2_\epsilon = \frac{\hat{\sigma}_f^2}{\gamma}$.
    \end{algorithmic} 
\end{algorithm}
\begin{algorithm}
    \caption{Scale the mean function}
    \label{alg:scale_mean}
    \begin{algorithmic}[1]
    \State Given a model of the outcome $Y = f(X) + \epsilon$ with independent, additive Gaussian noise $\epsilon \sim N(0, \sigma_\epsilon^2)$, the predictors $X \sim N(0,\Sigma_x)$, a fixed noise variance $\sigma_\epsilon^2$, and the desired SNR $\gamma$.
    \State Sample $x_i \sim N(0,\Sigma_x), i=1,...,m$ for large $m$.
    \State Estimate $\hat{\sigma}_f^2 = \frac{1}{m-1}\sum_{i=1}^m (f(x_i) - \frac{1}{m}\sum f(x_i))^2$.
    \State Calculate a scaling factor $ s = \gamma \frac{ \sigma_\epsilon^2}{\hat{\sigma}_f^2}$.
    \State Set $f_s = \sqrt{s} f$.
    \end{algorithmic}
\end{algorithm}

\subsection{Statistical methods for environmental mixtures}

Our package interfaces with several existing and newly developed analysis strategies for assessing associations between correlated exposures and health outcomes. These strategies can be used as inference models for Algorithm~\ref{alg:power_mc}. We provide an overview of methods and their unique strengths. We do not go into the mathematical description, and instead give intuitions that help researchers select the correct model given characteristics of the data. For a more detailed review of recent methodological developments, see \cite{ijerph19031378}.

\subsubsection*{Bayesian Model Averaging}

Model selection is a common strategy to deal with multicollinearity or large dimensionality in the predictors. Instead of selecting for one best model that might be sensitive to the given observed data set, BMA provides a posterior over the space of all possible models \citep{hoetingbma}. Thus, BMA allows for direct model selection and robust effect estimation by averaging the posteriors of the parameters under all models weighted by their posterior model probabilities. Unlike other model selection methods (i.e. forward/backward selection), BMA quantifies uncertainty more fully, including model uncertainty. Interactions and nonlinear effects can be included by transformation of the design matrix. Additionally, BMA can be used with many models, such as linear regression, generalized linear models, and survival models. Our package integrates the implementation of BMA in package \pkg{BMA} \citep{bmapackage}. As the dimension of the model space grows exponentially with the number of predictors, for high-dimensional data sets, implementations of BMA often approximate the model posterior by doing preliminary step-wise model selection (e.g. in package \pkg{BMA}) or by sampling the model space. In the presence of high correlations among many predictors, BMA is prone to produce low marginal inclusion probabilities for the correlated predictors, making it difficult to identify critical predictors of the outcome.

\subsubsection*{Bayesian Kernel Machine Regression}

Bayesian kernel machine regression can flexibly model nonlinear, high-order interaction effects of multiple predictors jointly using a kernel function \citep{bkmr}. BKMR incorporates component-wise variable selection, producing a posterior inclusion probability for each individual variable, allowing for identification of critical components. Similar to BMA, BKMR is prone to produce low posterior inclusion probabilities for a large number of highly correlated variables even when many or all of them are important. To alleviate this issue, BKMR also implements hierarchical variable selection, which selects on groups of highly correlated variables. Researchers will need to specify the groups based on domain knowledge of the data set. BKMR can be prone to overfitting \citep{mixselect}. Our package integrates the implementation of BKMR from the package \pkg{bkmr} available on CRAN \citep{bkmr-package}.

\subsubsection*{MixSelect}

MixSelect decomposes the fitted mean function of the outcome into linear main effects, pairwise interactions, and a nonlinear deviation \citep{mixselect}. MS also include variable selection on the linear main effects and hierarchical variable selection on the interactions. This means that a pairwise interaction is included only if the main effects of one or both variables are selected. The nonlinear deviation models the residuals of the linear terms, using a kernel function similar to BKMR. Unlike BKMR, MS uses Bayesian model selection to decide whether to include the nonlinear deviation in the model. Thus, it automatically simplifies to a linear model with variable selection if only linear effects are present in the data. While BKMR relies solely on visualizations for interpretation of marginal effects, MS produces posteriors for the linear terms from which the mean, credible intervals, and other statistics can be calculated. However, MS does not offer group-wise variable selection like BKMR. Like BKMR, MS suffers with data of large dimensionality, sample size, and extensive multicollinearity. We use the implementation provided by the authors at \url{https://github.com/fedfer/MixSelect}.

\subsubsection*{Factor Analysis for Interactions}

Factor analysis is a common choice to deal with multicollinearity in the predictors. It can infer group structures in the predictors, i.e. latent factors, while being informed by their relationships to the outcome, effectively reducing the dimension of the data. \citep{fin} proposed the FIN model that accounts for pairwise interactions between the factors and transforms the parameters to produce linear main effects, quadratic main effects and pairwise interactions in the original predictors. Thus, it examines effects of combinations of predictors as well as individual predictor at a reduced model complexity. Reasonable assumptions about group structures in the predictors must hold. FIN does not produce exact variable selection, though it induces shrinkage on the effect estimates. FIN might not be effective at identifying individual critical predictors, especially when other correlated predictors are not important. We use the implementation provided in the package \pkg{infinitefactor} \citep{finpackage} available on CRAN.

\subsubsection*{Bayesian Weighted Sums}

Bayesian Weighted Sums aims at estimating a combined effect and individual variable importance by summing multiple predictors into a single exposure variable using quantitative weights \citep{bws}. Thus, BWS reduces the dimension of the problem and produces a parsimoneous model. BWS may be thought of as a factor model that assumes only one latent factor affects the outcome. The weights can be interpreted as proportions of the combined effect explained by each component. Unlike QGC (discussed next), BWS provides uncertainty quantification on the weights. BWS is currently limited to modeling linear additive effects and constrains all components of the mixtures to have effects of the same sign. We use the implementation of BWS in the package \pkg{bws} \citep{bwspackage} available on CRAN.

\subsubsection*{Quantile G-Computation}

Similar to BWS, QGC aims to estimate the combined effects of multiple predictors. QGC first transforms the predictors into categorical variables whose categories correspond to quantiles of the original predictors. QGC then fits a generalized linear model to the categorical variables, and the sum of all regression coefficients produces the g-computation estimator. Since the underlying model is a generalized linear model, if these three assumptions hold: 1) the relationship is in fact linear, 2) quantization of predictors is appropriate, and 3) there are no unmeasured confounders, then the g-computation estimator gives a causal dose-response parameter \citep{qgcomp}. Unlike BWS, QGC allows predictors to have effects of different signs and allows the inclusion of nonlinear and interaction terms. Identification of critical components may be difficult because the weights lack confidence intervals or p-values. We use the \texttt{R} package \pkg{qgcomp} available on CRAN \citep{qgcomppackage}.

\section[Functionality]{Tutorial: Functionalities} \label{sec:tutorial}

We will walk through all functionality of the package. There are three data models that are building blocks to running a power analysis with this package: a generative model for the predictors, a generative model for the outcome given the predictors, and an inference model.

\subsection{Specifying a predictors model}

If sufficient existing data are available, simulated predictors may be created by resampling. For example, if we were planning a study of metabolic disease and phthalates, we could sample phthalate levels from NHANES data:

\begin{lstlisting}
R> xmod <- MixtureModel(method = `resampling', data = nhanes, 
+    resamp_prob = weights)
\end{lstlisting}

A vector of sampling probabilities of the observations, which must sum to 1, may be given to argument \code{resamp\_prob}. If limited data exist, predictors may be sampled from a semi-parametric Bayesian copula parametrized by the empirical univariate marginals and a correlation matrix \citep{hoff2007sgc}. The marginals and correlation matrix of the latent variables may be estimated directly as discussed in Section 2.2. This method can handle mixed-scaled data (i.e., continuous, binary, ordinal). The generated predictors are on the same scale as the original data. Note that categorical and ordinal data must be formatted as numeric (e.g. using one-hot-encoding). Users may pass a list of named arguments, including the number of MCMC samples (\code{nsamp}), how much to thin a chain (\code{odens}), and whether to print MCMC progress (\code{verb}) to the argument \code{sbg\_args}. See function \code{sbgcop.mcmc()} from package \pkg{sbgcop} for the full list of arguments.

\begin{lstlisting}
R> sbg_args <- list(nsamp = 2000, odens = 2, verb = TRUE) 
R> xmod <- MixtureModel(method = `estimation',
+    data = existing_data, sbg_args = sbg_args)
\end{lstlisting}

Alternatively, if no existing data are available to estimate the correlation matrix of the latent variables and univariate marginals, users can specify them manually using Algorithm \ref{alg:alg_rand_cor}. The argument \code{G} takes a $p \times p$ matrix of rough guesses of pairwise correlations between all variables. Larger values for the argument \code{m} corresponds to smaller deviation from \code{G} in the random samples. In the code below, we generate random matrices with high pairwise correlations around 0.9, as well as a random matrix with group structures:

\begin{lstlisting}
R> set.seed(1)
R> p <- 40
R> variable_names <- paste0(`x', 1:p)
R> G <- diag(1, p)
R> G[upper.tri(G)] <- 0.9
R> G[lower.tri(G)] <- 0.9
R> cvine_marginals <- lapply(1:p, function(i) `qnorm(mean = 0, sd = 1)')
R> colnames(G) <- rownames(G) <- names(cvine_marginals) <- variable_names
R> small_devi <- MixtureModel(method = `cvine', G = G, m = 100,  
+    cvine_marginals = cvine_marginals)
R> large_devi <- MixtureModel(method = `cvine', G = G, m = 10, 
+    cvine_marginals = cvine_marginals)
R> G_group <- matrix(c(rep(c(rep(0.8, p/2), rep(0.3, p/2)), p/2), 
+    rep(c(rep(0.3, p/2), rep(0.7, p/2)), p/2)), p, p)
R> diag(G_group) <- 1
R> group <- MixtureModel(method = `cvine', G = G_group, m = 100,
+    cvine_marginals = cvine_marginals)
\end{lstlisting}

\begin{figure}[!ht]
\centering
\includegraphics[width=0.7\linewidth]{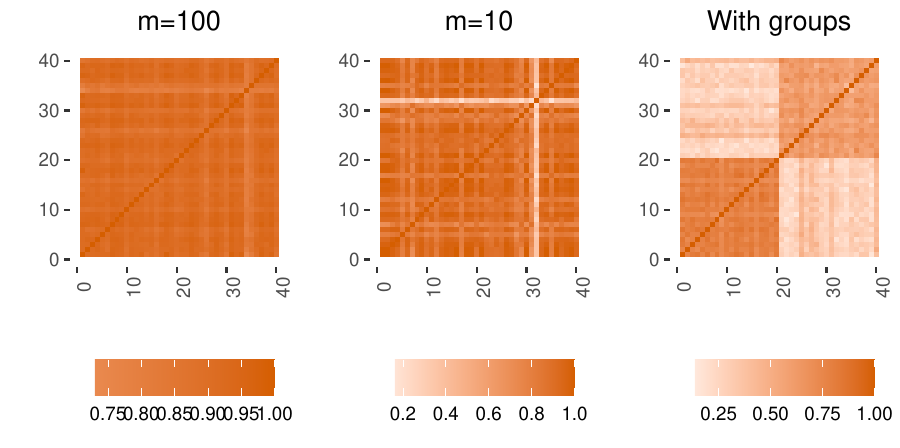}
\caption{Examples of random correlation matrices generated using Algorithm \ref{alg:alg_rand_cor}.}
\label{fig:mS}
\end{figure}

The sampled correlations concentrate more around 0.9 when \code{m}=100 than when \code{m}=10 (Figure~\ref{fig:mS} left and Figure~\ref{fig:mS} middle). The right-most subplot of Figure~\ref{fig:mS} shows an example matrix created by an input \code{G} with group structures. The argument \code{cvine\_marginals} takes a named list of strings describing the quantile function and named parameters of the marginal distribution of each variable. Valid strings include standard distributions from the \texttt{R} package \pkg{stats}. See our documentation for a complete list of supported distributions. For an example of how to create independent binary, count, and ordinal variables, see the first example in Section \ref{sec:example}.

%

\subsection{Specifying an outcome model}
Users need to specify a hypothesized model for the outcome given the predictors. This can be thought of as the ``true'' relationship between the predictors and outcome. This model may be informed from prior knowledge or pilot studies. When little information is available, investigators may wish to test a variety of outcome models to get a sense of robustness to assumptions about predictor-outcome relationships. The model for the mean is given as a string of \texttt{R} code. For example, if the continuous outcomes are modeled as random draws from a Gaussian $N(2(x_1 + x_2), 1)$, we can use the following code:

\begin{lstlisting}
R> ymod <- OutcomeModel(f = `2*(x1 + x2)', family = `gaussian', sigma = 1)
\end{lstlisting}

The function can generate binary and count data using an appropriate link function (i.e., logit) when \code{family} is set to \code{`binomial'} or \code{`poisson'} respectively. A fitted model of the outcome to existing data may also be used but needs to be wrapped inside a function that takes a predictor matrix \code{|x|} and returns a vector of mean values. Below is an example of wrapping an \code{lm} model to be used with our package:

\begin{lstlisting}
R> pilot_mod <- lm(y ~ x1, data = pilot_study)
R> pilot_f <- function(x) {
+      predict(pilot_mod, as.data.frame(x))$fit
+  }
R> pilot_sigma <- sigma(pilot_mod)
R> ymod <- OutcomeModel(f = pilot_f, family = `gaussian', 
+      sigma = pilot_sigma)
\end{lstlisting}

\subsection{Specifying an inference model}
While the outcome and the inference models are ideally the same, in practice our model will at best approximate the ``true'' predictor-response relationships. In the code below, we show how to use BKMR as the inference model. The \verb|model| argument also accepts one of the following values: \verb|`glm'|, \verb|`mixselect'|, \verb|`qgcomp'|, \verb|`bws'|, \verb|`fin'|, and \verb|`bma'|. Advanced users may also define custom inference models (see first example in Section \ref{sec:example}). Additional arguments specific to the inference model (e.g. \code{iter}, \code{varsel}) can be passed as done below:

\begin{lstlisting}
R> imod <- InferenceModel(model = `bkmr', iter = 10000, varsel = TRUE)
\end{lstlisting}

The built-in inference functions return values, such as p-value or posterior inclusion probability, as ``significance'' criteria. Table~\ref{tab:sig-criteria} list the criteria available for each built-in inference model.

\begin{table}[ht]
    \centering
    \begin{tabular}{| p{0.1\linewidth} | p{0.55\linewidth}| p{0.25\linewidth}|}
        \hline
        Inference method & Individual effect & Value to \verb|crit| \\
        \hline
        BMA & posterior inclusion probability (PIP) of a main effect or an interaction term $\geq \alpha$. & \verb|`pip'|\\ 
        \hline
        GLM & the smallest p-value of the regression coefficients (i.e., main effect, interactions) involving a predictor $\leq \alpha$. & \verb|`pval'|, \verb|`main_pval'| (main effect only), \verb|`int_pval'| (interactions only)\\
        \hline
        BKMR & PIP of a non-zero length-scale parameter $\geq \alpha$. & \verb|`pip'| (component-wise), \verb|`group_pip'| (group-wise)\\
        \hline
        MS & PIP of a main effect or a length-scale parameter $\geq \alpha$. & \verb|`pip'|, \verb|`linear_pip'| (main effect alone), \verb|`gp_pip'| (lenghth-scale alone)\\
        \hline
        FIN & $(1-\alpha)$ \% CI of a main effect or an interaction term doesn't include zero. Equivalent to $min(Pr(\beta >0|.), Pr(\beta < 0|.)) \leq \alpha$ where $Pr(\beta < 0|.)$ is the posterior probability of an effect being less than 0. & \verb|`beta'|, \verb|`linear_beta'| (main effect alone)\\ 
        \hline
        BWS & $(1-\alpha)$ \% CI of the joint effect doesn't include zero. Equivalent to $min(Pr(\beta >0|.), Pr(\beta < 0|.)) \leq \alpha$ where $Pr(\beta < 0|.)$ is the posterior probability of the joint effect being less than 0. & \verb|`beta'|\\ 
        \hline
        QGC & p-value of the joint effect $\leq \alpha$. & \verb|`pval'|\\
        \hline
    \end{tabular}
    \caption{Definitions of effect criteria for built-in inference models given a threshold $\alpha$.}
    \label{tab:sig-criteria}
\end{table}

\subsection{Running a power analysis}
Function \code{sim\_power()} executes steps 2 to 6 of Algorithm \ref{alg:power_mc} given a predictors generative model, an outcome generative model, an inference model, a sample size, and the number of MC iterations. In the code below, we simulate 2000 data points specified by the predictors and outcome model, fit an inference model and repeat the whole process 10000 times:

\begin{lstlisting}
R> res <- sim_power(xmod, ymod, imod, s = 10000, n = 2000, 
+    snr_iter = 100000, cores = 2, errorhandling = `stop', 
+    cluster_export = c(`pilot_mod'))
\end{lstlisting}

Each simulation is independent and thus may be parallelized for faster computation. When multiple cores are available on the computer, increase \verb|cores| to parallelize the simulation. Internally, we use the \pkg{doSNOW} \citep{dosnow} and \pkg{foreach} \citep{foreachpackage} packages for this feature. While using parallelism, if an error occurs in any iteration, \verb|errorhandling| specifies whether to remove that iteration (\verb|`remove'|), return the error message verbatim in the output (\verb|`pass'|), or terminate the loop (\verb|`stop'|). The \verb|`pass'| and \verb|`stop'| options are useful for debugging. Larger values for \verb|snr_iter| results in more precise estimates of the SNR. Finally, any global variable not explicitly passed in as an argument (e.g. a custom outcome model) needs to be exported to the parallel cluster using \code{cluster\_export}.

\subsection{Creating a power curve}

It is often desirable to study the trade-offs between power, Type I error rate, sample size, and effect size. Function \code{sim\_curve()} generates a power curve by running the power analysis described previously on a combination of sample sizes and outcome models. Different effect sizes may be baked into the outcome models and measured by the signal to noise ratio. Function \code{sim\_curve()} takes a list of outcome model and a vector of sample sizes, but only one predictors model and one inference model at a time. Below we conduct power analyses for 9 combinations of sample sizes 1000, 2000, and 5000 and 3 outcome generative models:

\begin{lstlisting}
R> res_curve <- sim_curve(xmod = xmod, imod = imod, 
+    ymod = list(ymod1, ymod2, ymod3), s = 10000, n = c(1000, 2000, 5000))
\end{lstlisting}

\subsection{Estimating the signal-to-noise ratio}

As we discuss in Section \ref{sec:method}, the SNR is a good proxy for a generalization of an effect size to non-linear outcome model. The SNR can be viewed as an estimate for the ratio between the variation in the outcome explained by the true predictors and the variation due to noise. To estimate the SNR of a data generating process and its bootstrap standard error (s.e.) , we can pass a predictors model and an outcome model to the following function:

\begin{lstlisting}
R> estimate_snr(xmod, ymod, m = 100000, R = 1000)
\end{lstlisting}

Argument \verb|m| define the number of samples drawn from the predictors and outcome models to estimate SNR and \verb|R| the number of bootstrap replicates. See the Appendix \ref{app:secA1} for an illustration of how the estimated SNR changes as \code{m} varies.

If the outcome generative model has Gaussian white noise, we can automatically scale the mean function (use \code{scale\_f()}) or the noise variance (use \code{scale\_sigma()}) to get a desired SNR as discussed in Algorithms \ref{alg:scale_noise} and \ref{alg:scale_mean}. The code below modifies a given outcome model and returns a new one with the specified SNR:

\begin{lstlisting}
R> desired_snr <- 0.5
R> new_ymod <- scale_f(desired_snr, cur_ymod, xmod)
R> new_ymod <- scale_sigma(desired_snr, cur_ymod, xmod)
\end{lstlisting}

\subsection{Getting the summary}

The \code{summary()} function produces summaries statistics and plots for the power analysis. It takes a `\code{Sim}' object or a `\code{SimCurve}' object returned by \code{sim\_power()} or \code{sim\_curve()}. The \verb|crit| argument specifies the significance criterion; the \verb|thres| argument specifies the significance threshold; and \verb|how| describes how the criterion should be compared to the threshold (\verb|`lesser'| or \verb|`greater'|) . For example, for BMA, if PIP $> 0.5$ is considered ``significant'', then the criterion is PIP and the threshold is 0.5. For Bayesian parametric models, credible interval can also be used as a criterion, and p-values for frequentist models (e.g. QGC). Table~\ref{tab:sig-criteria} shows criteria available for each built-in inference model and corresponding values to give to \verb|crit|. The code below shows how to extract summaries for a BMA with PIP greater than 0.5, a GLM with p-values less than 0.01, and a BWS model with zero-coverage of the 90\% or larger credible interval as ``significance'' thresholds:

\begin{lstlisting}
R> summary(bma_res, crit = `pip', thres = 0.5, how = `greater')
R> summary(glm_res, crit = `pval', thres = 0.01, how = `lesser')
R> summary(bws_res, crit = `beta', thres = 0.1, how = `lesser')
\end{lstlisting}

Different thresholds will affect the power of the test. We may plot how the power changes for different thresholds:

\begin{lstlisting}
R> plot_summary(bma_res, crit = `pip', 
+    thres = c(0.5, 0.6, 0.7, 0.8), how = `greater')
\end{lstlisting}

\section[Synthetic Examples]{Tutorial: Examples} \label{sec:example}

In this section, we provide four examples of doing end-to-end power analysis with the functionality presented in the last section. The first two examples demonstrate that simulation provides precise estimates of power that match theoretical values. The first example involves using C-vine for simulating predictors and generating power curve for a general linear F-test. The second example involves simulating exposures from estimates of multivariate associations based on NHANES and again calculating power of a general linear F-test. The last two examples shows the importance of using the intended inference model for power analysis, especially in the presence of multicollinearity in the predictors. The third example shows a example of sample size planning for a mixtures study using closed-form formula compared to using MC simulation. The final example conducts power analyses for two mixtures methods that have different analysis goals.

\subsection{Manual specification of multivariate associations among predictors}

In the first example, we generate predictors using the C-vine method and calculate power for a general linear F-test. A closed-form power formula exists for the F-test and can be used to validate power estimated with MC simulations. The code below defines a generative model for two continuous, one binary, and one ordinal predictor with 3 levels, all independent of one another:

\begin{lstlisting}
R> xmod <- MixtureModel(method = `cvine', G = diag(1, 4),
+    cvine_marginals = list(x1 = `qnorm(mean=0, sd=1)',
+                           x2 = `qnorm(mean=0, sd=1)',
+                           x3 = `qbinom(size=1, prob=0.7)',
+                           x4 = `qmultinom(probs=c(0.5, 0.3, 0.2))'))
\end{lstlisting}

\begin{figure}[!ht]
\centering
\includegraphics[width=0.7\linewidth]{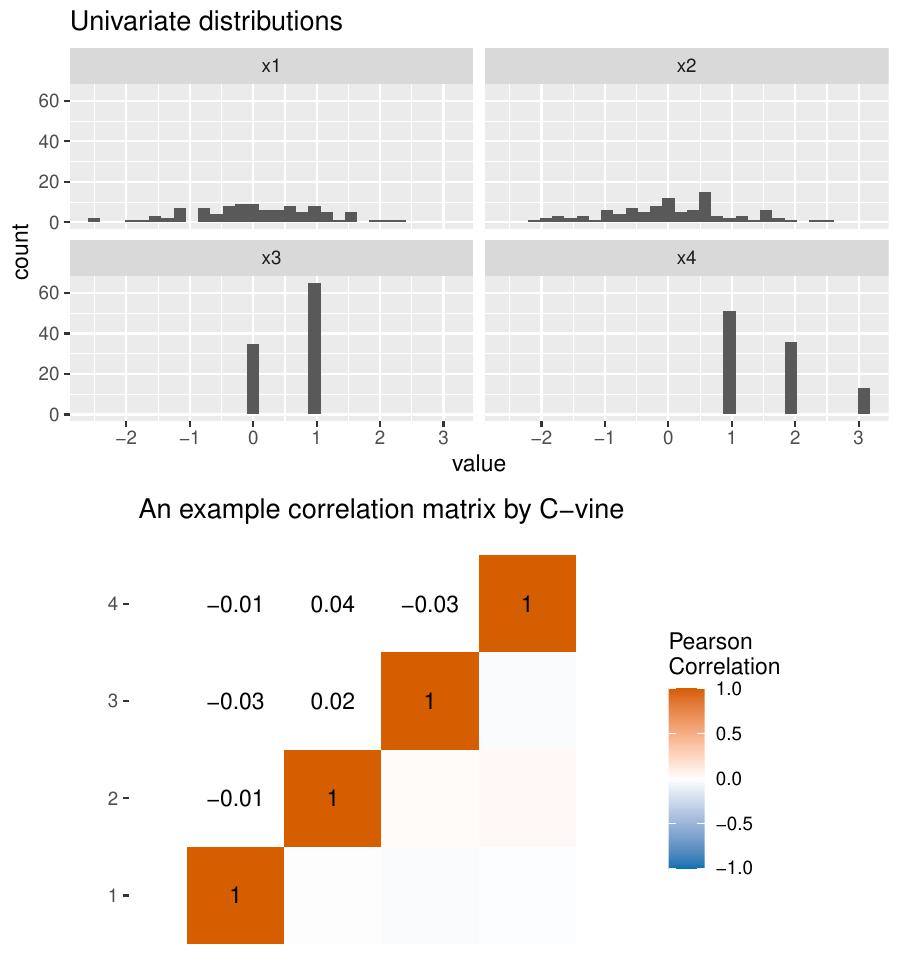}
\caption{Output of plotting the C-vine mixture model with \code{mplot(xmod)}. Note that the C-vine correlation matrix is for latent variables in Algorithm~\ref{alg:copula}.}
\label{fig:cvine-plot}
\end{figure}

Figure~\ref{fig:cvine-plot} shows the marginals and correlation matrix of the mixture model. Note that the C-vine correlation matrix is for latent variables in Algorithm~\ref{alg:copula}. We hypothesize the continuous outcome $y$ to be a linear function of $x_1$, $x_2$ and Gaussian white noise with variance 1. Mathematically, $y_i \sim N(\beta_1 x_{i1} + \beta_2 x_{i2}, 1)$ where $i$ indexes the observation, and $\beta_1, \beta_2$ are the regression coefficients. The code below defines three hypotheses on the outcome model with decreasing SNRs:

\begin{lstlisting}
R> ymod_list <- list(
+    OutcomeModel(f = `0.3 * x1 + 0.3 * x2', sigma = 1, family = `gaussian'),
+    OutcomeModel(f = `0.2 * x1 + 0.2 * x2', sigma = 1, family = `gaussian'),
+    OutcomeModel(f = `0.1 * x1 + 0.1 * x2', sigma = 1, family = `gaussian')
+  )
\end{lstlisting}

Since $x_1 \perp x_2$ and $x_1, x_2, y$ are Gaussian, the SNR can be calculated exactly analytically, e.g., $\frac{Var(0.3x_{1} + 0.3x_{2})}{\sigma^2} = \frac{0.3^2(1) + 0.3^2(1)}{1} = 0.18$ for the first, 0.08 for the second, and 0.02 for the third outcome model. Nevertheless, note that the empirical estimates of SNRs might deviate slightly from the truth, as discussed in the last section. Note also that in the ``true'' generative model, $x_3$ and $x_4$ are unrelated to $y$, but this is unknown to the analyst. Next, we define an inference model including all four predictors. Since the F-test is not a built-in inference model, we will need to define it in a way that is compatible with the \class{InferenceModel} object. Specifically, it needs  to take two arguments \code{x}, \code{y} for a predictors matrix and a vector of outcomes respectively, and to return a named list of significance criteria and values to be compared to a threshold $\alpha$ later. For the F-test, one significance criterion is the p-value from the overall F-test:

\begin{lstlisting}
R> ftest <- function(y, x) {
+    dat <- as.data.frame(cbind(y, x))
+    lm_mod <- lm(y ~ ., data = dat)
+    fstat <- summary.lm(lm_mod)$fstat
+    fpval <- pf(fstat[1], fstat[2], fstat[3], lower.tail = F)
+    names(fpval) <- `F-test'
+    return(list(pval = fpval))
+  }
R> imod <- InferenceModel(model = ftest, name = `F-test')
\end{lstlisting}

We now have all the components to run power analysis for sample sizes 50, 100, 150, and 200. Since we consider 4 sample sizes and 3 hypothesized outcome models, we will use function \code{sim\_curve()} to estimate 12 power calculations using 1000 MC simulations each:

\begin{lstlisting}
R> set.seed(1)
R> curve <- sim_curve(xmod, ymod_list, imod, s = 1000, 
+    n = c(50, 100, 150, 200), cores = 1) 
\end{lstlisting}

We can extract the estimated power of the ovarll F-test based on the significance criterion ``p-value $< 0.05$'' as a data frame with \code{summary()}:

\begin{lstlisting}
R> curve_df <- summary(curve, crit = `pval', thres = 0.05, how = `lesser')
\end{lstlisting}

%

Figure~\ref{fig:syn-1} shows that the power curve estimated with MC simulations closely resembles the analytical power calculations:

\begin{figure}[!ht]
\centering
\includegraphics[width=0.7\linewidth]{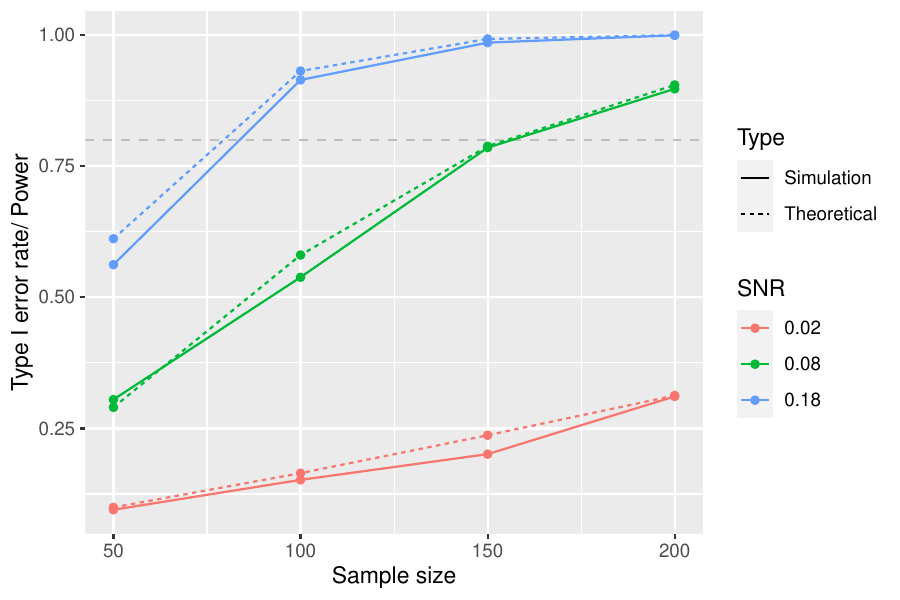}
\caption{Comparison of power curves for F-test using closed-form formula (theoretical) and MC simulations in C-vine example.}
\label{fig:syn-1}
\end{figure}

\subsection{Estimation of mixtures in NHANES}

\begin{figure}[!ht]
\centering
\includegraphics[width=0.9\linewidth]{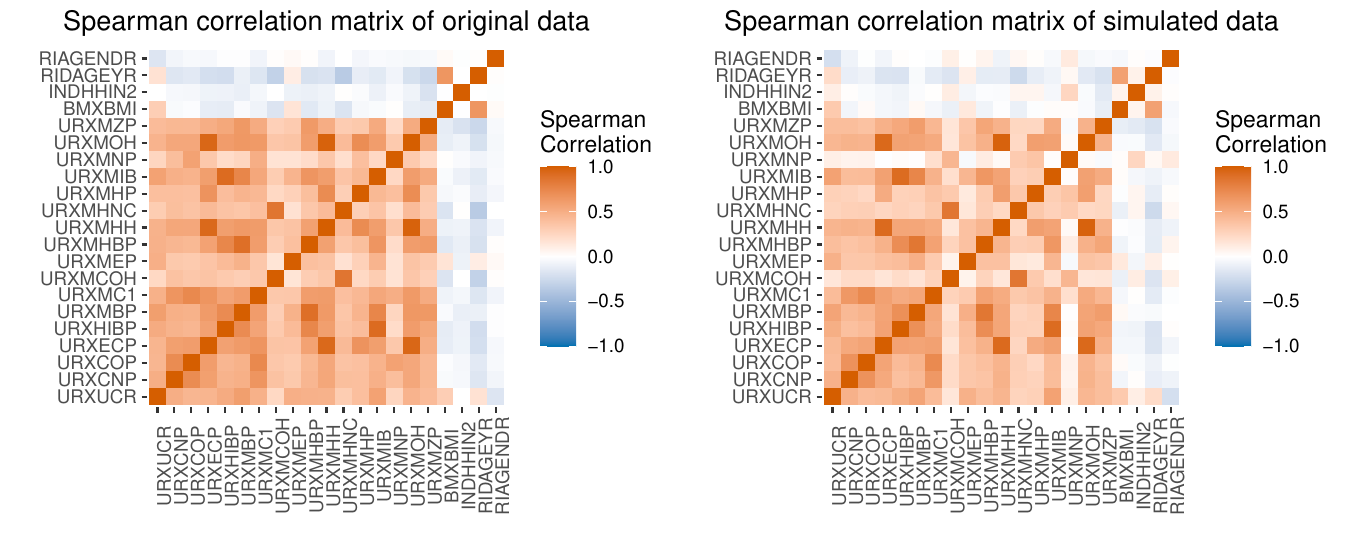}
\caption{Correlation matrices of mixture data from the NHANES 2015-2018 cycles (left) and of 500 observations simulated from an estimation mixture model (right).}
\label{fig:resamp-cor}
\end{figure}

In the second example, we generate predictors by estimating multivariate associations in NHANES data. We download publicly available demographic, examination and laboratory data from (\url{https://www.cdc.gov/nchs/nhanes.htm}). We combine data from the 2015-2016 and 2017-2018 cycles and adjust the survey weights as instructed in the official documentation (\url{https://wwwn.cdc.gov/nchs/nhanes/tutorials/module3.aspx}). The data is included in our package as \code{`nhanes1518'}. We use a data subset that includes 5483 observations with no missing data and the following variables: age (\code{RIDAGEYR}), gender (\code{RIAGENDR}), household income (\code{INDHHIN2}), BMI (\code{BMXBMI}), Creatinine (\code{URXUCR}), and 16 phthalate metabolites. We choose this data subset because it has continuous, binary, and ordinal predictors and has extensive correlations between many of them. We log-transform the phthalate metabolites and Creatinine. We use the Bayesian Gaussian copula to simulate synthetic data that preserves the dependence structure among the predictors. Figure~\ref{fig:resamp-cor} shows similar Spearman's correlation matrices between the original data (left) and a simulated data set of 500 observations (right):

\begin{lstlisting}
R> set.seed(1)
R> data(`nhanes1518', package = `mpower')
R> nhanes <- nhanes1518 %>%
+    select(starts_with(`URX'), BMXBMI, INDHHIN2, RIDAGEYR, RIAGENDR) %>%
+    filter(complete.cases(.)) %>%
+    mutate(across(starts_with(`URX'), log))
R> xmod <- MixtureModel(method = `estimation' , data = nhanes,
+    sbg_args = list(nsamp = 2000))
\end{lstlisting}

We then define the mean function for a continuous outcome as a linear combination between age, gender (male), Creatinine, and three correlated phthalate metabolites:

\begin{lstlisting}
R> ymod <- OutcomeModel(sigma = 1, family = `gaussian', 
+    f = `0.02*RIDAGEYR + 0.1*I(RIAGENDR==1) + 0.1*URXUCR + 
+    0.15*URXMHH + 0.07*URXMOH - 0.1*URXMHP')
\end{lstlisting}

We can use the \code{I()} function to define effects of different levels in a categorical variable. We consider the F-test defined in the first example again for the power analysis using 1000 MC simulations for a sample size of 100:

\begin{lstlisting}
R> set.seed(1)
R> power <- sim_power(xmod, ymod, imod, s = 1000, n = 100,
+    cores = 1, errorhandling = `stop')
\end{lstlisting}

The estimated SNR is between 0.27 and 0.28. For linear regression, the SNR is equivalent to the Cohen's $f^2$ effect size of all predictors. The closed-form formula for the general linear F-test at the 0.05 significance level with 21 predictors, 100 observations and an $f^2$ between 0.27 and 0.28 returns a power between 81\% to 85\%. At the same significance level, the MC estimate is:

\begin{lstlisting}
R> summary_df <- summary(power, crit = `pval', thres = 0.05, how = `lesser')
\end{lstlisting}

\begin{lstlisting}
	*** POWER ANALYSIS SUMMARY ***
Number of Monte Carlo simulations: 1000
Number of observations in each simulation: 100
Estimated SNR: 0.28
Inference model: F-test
Significance criterion: pval

Significance threshold:  0.05

|       | power|
|:------|-----:|
|F-test | 0.822|
\end{lstlisting}

%

We see from the example above that MC simulation is a reliable way to estimate power of a test.

\subsection{Sample size planning for a study to identify critical chemicals}

In the prior two examples, we demonstrate scenarios where both the closed-form formula and MC simulations produce nearly identical estimates. In this third example, we present a setting where closed-form formulae that don't account for the multicollinearity among predictors can produce inaccurate power estimates. Specifically, we show that sample size planning using simulation is advantageous compared to closed-form formulae when the study's objective is to identify individual critical chemicals from a correlated group. First, we simulate 11 phthalates exposures from NHANES by resampling the original data:

\begin{lstlisting}
R> chems <- c(`URXUCR', `URXMEP', `URXMBP', `URXMIB', `URXMHP', 
+      `URXMOH', `URXMHH', `URXECP', `URXMZP', `URXCOP', `URXMC1')
R> nhanes <- nhanes1518 %>% select(all_of(chems))
R> xmod <- MixtureModel(method = `resampling' , data = nhanes)
\end{lstlisting}

We define a synthetic outcome that is normally distributed around a linear function of log MEHHP (variable \code{URXMHH} in the \code{nhanes1518} data table). Mathematically, $y_i \sim N(1.5x_{i, log(MEHHP)}, 1)$. The code below defines the generative model for this outcome:

\begin{lstlisting}
R> ymod <- OutcomeModel(
+      f = `0.15*log(URXMHH)',
+      sigma = 1, family = `gaussian')
\end{lstlisting}

The true regression coefficients and important metabolites are again unknown to the researcher. The researcher is interested in designing a study to identify individual chemical that has an effect on the outcome. She plans to use a Bayesian linear model with variable selection (e.g. BMA) as the inference model. One naive approach to calculate the sample size for this study is to use the general linear F-test formula. Given that the full model is a multiple linear regression with 11 metabolites, and the reduced model is a multiple linear model withou MEHHP, we can calculate the true effect size of MEHHP to be $f^2=0.02$ \citep{cohen2013statistical}. We calculate the sample size using function \code{WebPower::wp.regression()} in \code{R} \citep{webpower}. At the 0.01 significance level, the analytical formula indicates we need a sample size of 600 to achieve 80\% power in detecting a small effect $f^2=0.02$ of one metabolite, controlling for the other 10 metabolites. We compare this estimate with sample size estimates by MC simulation. To do this, we first define two \code{InferenceModel}s. One is a general linear F-test testing for the effect of MEHHP above and beyond the other 10 metabolites, and the other is a BMA:

\begin{lstlisting}
R> ftest <- function(y, x) {
+      dat <- as.data.frame(cbind(y, x))
+      full_lm <- lm(y ~ ., dat)
+      reduced_lm <- lm(y ~ . - URXMHH, dat)
+      return(list(pval = anova(reduced_lm, full_lm)[[6]][2]))
+  }
R> imod_f <- InferenceModel(model = ftest, glm.family = `gaussian')
R> imod_bma <- InferenceModel(model = `bma', glm.family = `gaussian')
\end{lstlisting}

The BMA model will regress the outcome on all 11 chemicals simultaneously. We then generate power curves for sample sizes 600, 1000, 3000, and 5736 (since the original data has 5736 observations):

\begin{lstlisting}
R> set.seed(1)
R> results_bma <- sim_curve(xmod, ymod, imod_bma, s = 100, 
+      n = c(600, 1000, 3000, 5736),
+      cores = 3, errorhandling = `stop', snr_iter = 5000)
R> results_f <- sim_curve(xmod, ymod, imod_ftest, s = 100, 
+      n = c(600, 1000, 3000, 5736),
+      cores = 3, errorhandling = "stop", snr_iter = 5000)
\end{lstlisting}

To plot the power curve to detect the effect of MEHHP above and beyond the effects of the other 10 metabolites at the 0.01 significance level, we use the following code:

\begin{lstlisting}
R> plot_summary(results_f, crit = `pval', thres = 0.01, 
+      how = `lesser', digits = 3)
\end{lstlisting}

To plot power curves for the posterior inclusion probability of each metabolite being higher than 75\%, we use the following code:

\begin{lstlisting}
R> plot_summary(results_bma, crit = `pip', thres = 75, 
+      how=`greater', digits = 3)
\end{lstlisting}

\begin{figure}[!ht]
\centering
\includegraphics[width=0.97\linewidth]{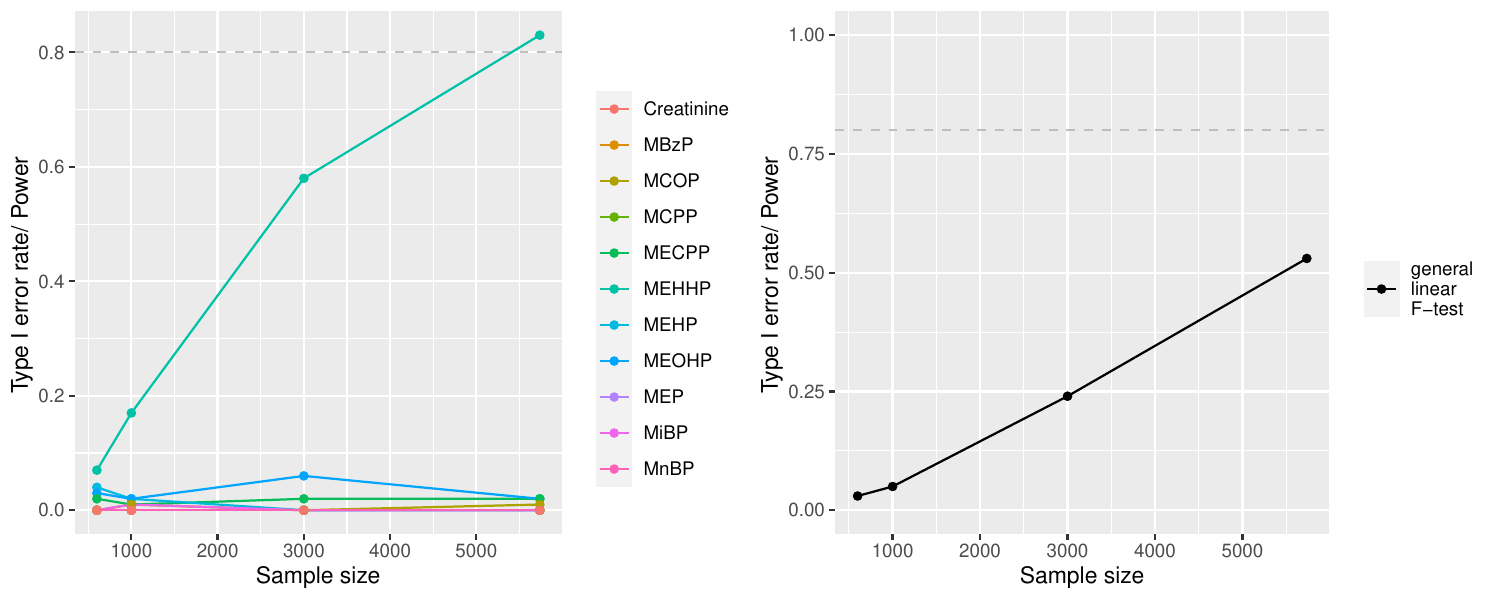}
\caption{Left: Power curve of the posterior inclusion probability of MEHHP in a Bayesian linear model averaging being higher than 75\%, while controlling for other metabolites. Lines for metabolites other than MEHHP show Type I error curves. Right: Power curve to detect a small effect of MEHHP above and beyond the effects of the other 10 metabolites at the 0.01 significance level.}
\label{fig:thres-syn2}
\end{figure}

In the left side of Figure \ref{fig:thres-syn2}, the MEHHP line shows that BMA has less than 10\% power to identify MEHHP as an important chemical with 600 study subjects, controlling for the other metabolites, given that 1 unit increase in log MEHHP is associated with 0.15 unit increase in the outcome. As the number of subjects increases to approximately 5500, BMA's power increases to 80\%. These results suggest that almost 60000 observations are needed to detect individual critical chemicals with at least 80\% power. The lines for all other chemicals (Creatinine, MBzP, etc.) show the Type I error rate (proportion of rejected tests among all simulations when there is actually zero effect) as a function of sample size. 

In the right side of Figure \ref{fig:thres-syn2}, we see that the general linear F-test has close to 0\% power to detect the small effect of MEHHP while controlling for other metabolites with 600 subjects. This is considerably lower than the estimate provided by the analytical formula. This discrepancy arises because the chemical mixtures involves multicollinearity, which inflates the variances in unregularized multiple linear regressions. Additionally, we see that BMA is more powerful than the F-test at identifying individual important chemicals for a fixed sample size.

\subsection{An example in reproducing a study}

A recent study \citep{wu2020nhanes} considers the relationships between mixed exposures to various chemicals and obesity in children and adolescents using NHANES 2005-2010 data. Obesity in the study is measured by BMI z-scores and an indicator of BMI z-scores being over the $95^{th}$ percentile. All exposures are log transformed for the analysis. The results from various statistical models suggest that Dichlorophenol25 and MEP may be important risk factors for obesity. Suppose that we want to design a study to replicate these results, possibly in a specific population of interest. An important first step is to calculate a sample size to ensure that the study is well-powered and that we avoid dedicating resources to collecting more data than needed. We may specify the target power to be 80\% and the target Type I error rate to be $0.05$, which are common practices. We also need to specify effect sizes appropriately. One may simply use the estimated effects of 2,5-DCP and MEP from the original study using the NHANES 2005-2010 data provided by the authors. Specifically, the original study finds one unit increase in Dichlorophenol25 (on log scale) is associated with a $e^{0.55} = 1.73$ times increase and MEP (on log scale) with a $e^{0.31} = 1.36$ times increase in the odds of obesity in an multivariate logistic regression adjusted for all covariates and chemicals \citep[see Additional file 1 Table S1]{wu2020nhanes}. We additionally hypothesize that there may be a modest interaction between MEP and Methylparaben. Thus, we will use the following conditional mean function for the binary response obesity:

$$ logit(Pr(y_i=1)) = 0.55x_{i,log(dichlorophenol25)} + 0.31x_{i, log(MEP)} + 0.2x_{i, log(MEP)}x_{i, log(Methylparaben)} $$

We conduct power analysis for three sample sizes $200, 500$, and $1000$ using both a logistic regression as well as BWS as inference models. We can summarize the power at each sample size by evaluating the p-values at the 0.05 significance level for each predictor separately in the logistic regression. Likewise, we look at the $95\%$ credible interval of the joint effect in the BWS model. The following code demonstrates how to do that with our package:

\begin{lstlisting}
R> set.seed(1)
R> data_url <- paste0(`https://static-content.springer.com/esm/',
+                     `art%3A10.1186%2Fs12940-020-00642-6/MediaObjects/',
+                     `12940_2020_642_MOESM2_ESM.xlsx')
R> nhanes <- openxlsx::read.xlsx(data_url)
R> chems <- c(`UrinaryBisphenolA', `UrinaryBenzophenone3',
+             `Methylparaben', `Propylparaben',
+             `dichlorophenol25', `dichlorophenol24',
+             `MBzP', `MEP', `MiBP')
R> xmod <- MixtureModel(data = nhanes[, chems], method = `resampling')
R> ymod <- OutcomeModel(
+    f = `0.55*dichlorophenol25 + 0.31*MEP + 0.2*MEP*Methylparaben', 
+    family = `binomial')
R> bws_logit <- InferenceModel(model = `bws', iter = 5000, chains = 1,
+    refresh = 0, family = `binomial')
R> glm_logit <- InferenceModel(model = `glm', family = `binomial')
R> n_cores <- 1
R> s <- 100
R> n <- c(200, 500, 1000)
R> bws_out <- sim_curve(xmod = xmod, ymod = ymod, imod = bws_logit, 
+    s = s, n = n, cores = n_cores)
R> glm_out <- sim_curve(xmod = xmod, ymod = ymod, imod = glm_logit, 
+    s = s, n = n, cores = n_cores)
\end{lstlisting}

\begin{figure}[!ht]
\centering
\includegraphics[width=0.9\linewidth]{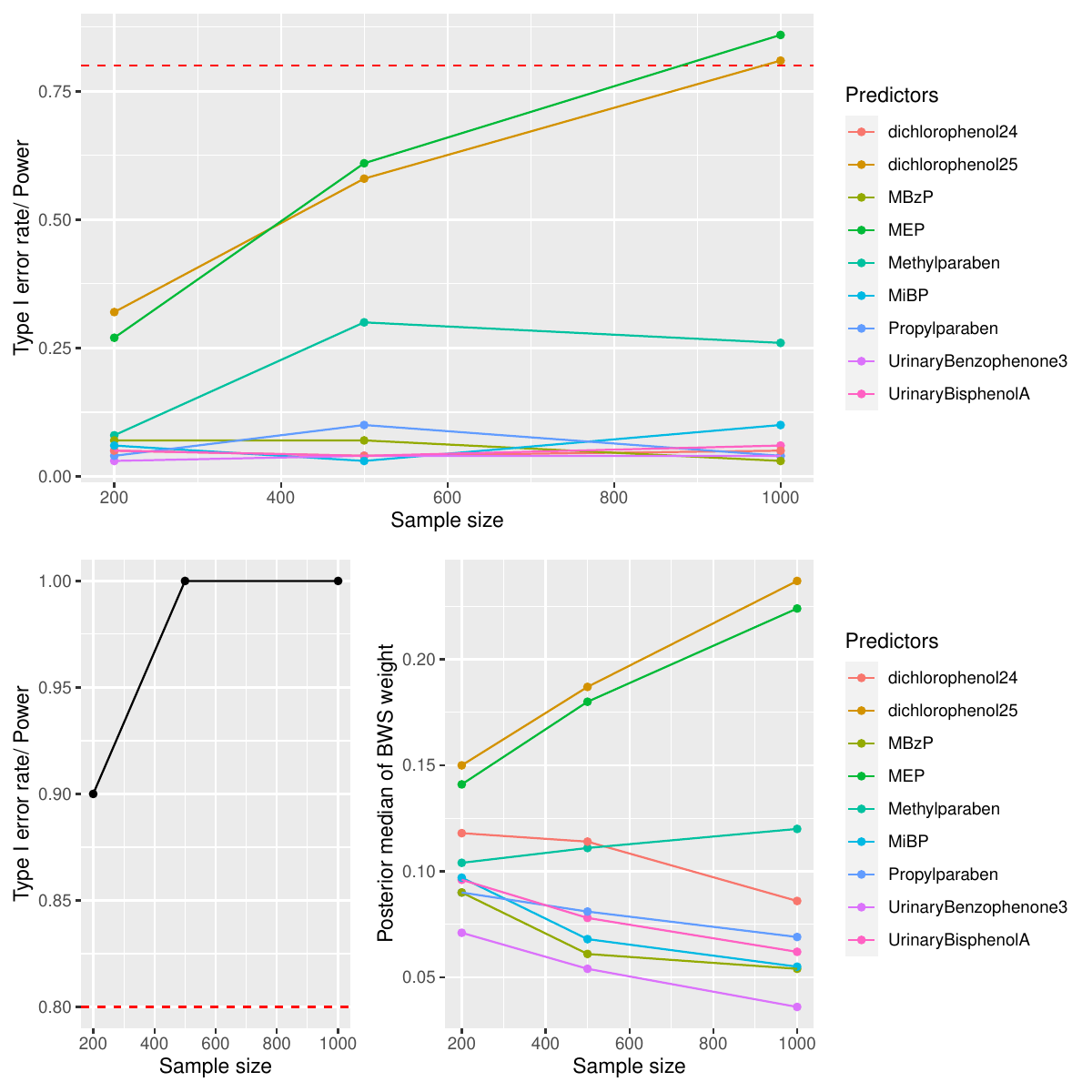}
\caption{Top: Power curves for Dichlorophenol25, MEP, and Methylparaben from a logistic regression at a 0.05 significance level. The other curves are Type I error rate curves for chemicals with null effects. Bottom left: Power curve of the joint effect from a BWS using a $95\%$ credible interval. Bottom right: The posterior median BWS weights of the predictors. In all subplots, the dotted red line represents the popular 80\% threshold for having adequate power.}
\label{fig:nhanes-example}
\end{figure}

The top subplot in Figure~\ref{fig:nhanes-example} shows estimated power curves to detect individual effects in a multiple logistic regression for Dichlorophenol25, MEP, and Methylparaben. The  Dichlorophenol25 line shows that we need around 1000 study subjects to achieve 80\% power to detect an odds-ratio of 1.73 between subjects that differ by 1 unit of this chemical, controlling for all other chemicals in a logistic regression. We need a sample considerately larger than 1000 to detect the interaction between MEP and Methylparaben. The curves for chemicals with no true effect on the outcome indicate Type I error rate as a function of the sample size.

The subplot in the bottom left of Figure~\ref{fig:nhanes-example} shows that the power to detect the joint effect in BWS is much higher than the power to detect individual effects in a logistic regression for small samples. There is 90\% power to detect a joint effect, which represents the effect of a weighted sum of all chemicals, at a sample size of 200. The weights from BWS, however, could be misleading since they erroneously show Dichlorophenol24, a compound closely related to Dichlorophenol25, having a moderate contribution to the joint effect. However, this error diminishes as the sample size increases. Thus, if the researchers are mainly interested in the joint effect of a chemical mixtures, they may save resources by collecting fewer observations and using an inference model such as BWS or QGC. However, if individual effects are of interest, a large sample is needed, and logistic regression, BMA, or BKMR may be preferred.

\section{Conclusion}\label{sec:conclusion}

We provide an \texttt{R} package that allows researchers to quickly set up MC simulation for power analysis of observational studies of environmental exposure mixture. The package supports power analysis for recently developed statistical methods for mixtures that lack closed-form power formulas. It allows users to simulate realistic multivariate associations among exposures and mixed-scaled predictors using existing data. This is important to mixture studies because moderate to high correlations among the predictors can diminish the power of many statistical tests. Through our examples, we highlight the importance of conducting power calculations and sample size planning using the mixture method intended to be used for data analysis, while also emphasizing the advantages of simulation-based power analysis in exposure mixture studies.

\backmatter

\bmhead{Supplementary information}

\bmhead{Acknowledgments}

This work was partially supported by grants R01ES027498 and R01ES028804 of the National Institute of Environmental Health Sciences of the United States National Institutes of Health. 


\section*{Declarations}

\subsection*{Funding}

This work was partially supported by grants R01ES027498 and R01ES028804 of the National Institute of Environmental Health Sciences of the United States National Institutes of Health.

\subsection*{Competing interests}

The authors have no competing interests to declare that are relevant to the content of this article.

\subsection*{Ethics approval}

Not applicable

\subsection*{Consent to participate}

Not applicable

\subsection*{Consent for publication}

Not applicable

\subsection*{Availability of data and materials}

Publicly available on Github (\url{https://github.com/phuchonguyen/mpower}).

\subsection*{Code availability}

Publicly available on Github (\url{https://github.com/phuchonguyen/mpower}).

\subsection*{Authors' contributions}

Amy H. Herring devised and supervised the project. Phuc H. Nguyen developed the software, examples, and wrote the first draft of the manuscript. Stephanie M. Engel provided critical feedback and helped shape features of the software. All authors contributed to the final manuscript.

%
%
%
%

\begin{appendices}

\section{Estimated signal-to-noise ratio as a function of \code{m}}\label{app:secA1}

We will estimate the SNR of the following data-generating process using different values for \code{m}:

\begin{lstlisting}
R> set.seed(1)
R> xmod <- MixtureModel(method = `resampling', 
+    data = data.frame(x1 = rnorm(200000, mean = 0, sd = 1),
+                      x2 = rnorm(200000, mean = 0, sd = 1)))
R> ymod <- OutcomeModel(f = `0.3 * x1 + 0.3 * x2', 
+    sigma = 1, family = `gaussian')
R> for (m in c(500, 5000, 50000, 100000)){
+    estimate_snr(ymod, xmod, m = m, R = 1000)
+  }
\end{lstlisting}

\begin{figure}[!ht]
\centering
\includegraphics[width=0.7\linewidth]{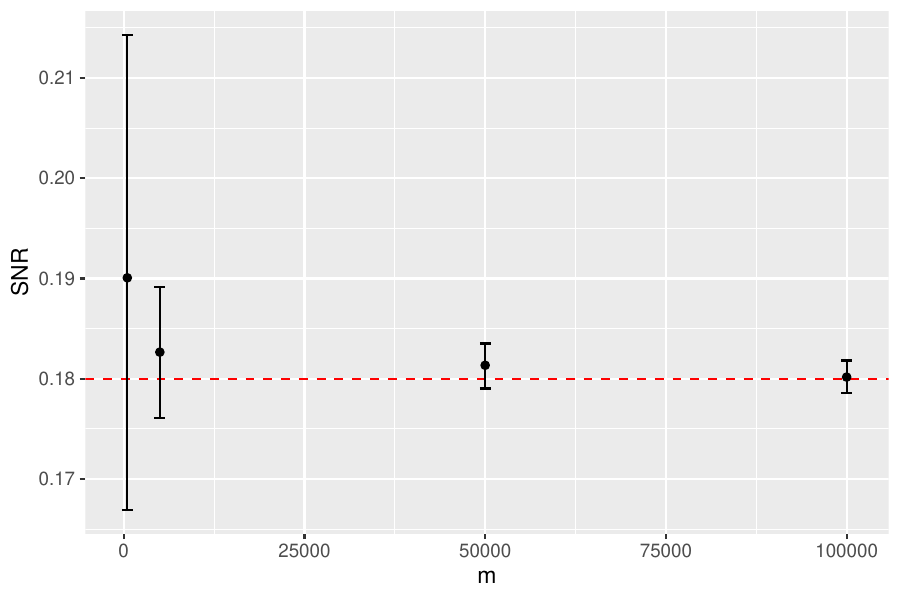}
\caption{The estimated SNR for the linear model example is unbiased but the standard error might be large with a small sample of simulated data. The red horizontal line is the ground truth SNR.}
\label{fig:snr-compare}
\end{figure}

Since the predictors are independent standard normal distributions, and the noise variance is 1, we can calculate the true SNR as $[0.3^2(1) + 0.3^2(1)]/1 = 0.18$. Figure~\ref{fig:snr-compare} shows the estimated SNR and 1000-bootstrap s.e. for \code{m} $\in \{500, 5000, 50000, 10000, 200000\}$. A larger \verb|m| results in a more precise estimate. When the mixture model is defined based on resampling, it may not be possible to choose a large \verb|m| without duplicating observations and underestimating the signal.

\end{appendices}


\bibliography{sn-bibliography}

\end{document}